\documentclass{PoS}

\usepackage{amsmath}
\usepackage{graphicx}
\usepackage{mathrsfs}

\title{Lattice Calculation of the $K_L$-$K_S$ mass difference}

\ShortTitle{Lattice Calculation of the $K_L$-$K_S$ mass difference}

\author{\speaker{Jianglei Yu}\\
       Department of Physics, Columbia University, New York, NY 10025, USA \\
       E-mail: \email{jy2379@columbia.edu}}

\abstract{
We report progress on calculating the $K_L$-$K_S$  mass difference in lattice QCD. The calculation is performed on a 2+1 flavor, domain wall fermion, $16^3\times 32$ ensemble with a 421 MeV pion mass. We include only current-current operators and drop all disconnected and double penguin diagrams in the calculation. The calculation is made finite through the GIM mechanism by introducing a valence charm quark. The long distance effects are discussed separately for each of the two parity channels. While we find a clear long distance contribution from the parity odd channel, the signal to noise ratio in the parity even channel is exponentially decreasing and the two-pion state can be seen in only a subclass of amplitudes. We obtain the mass difference $\Delta M_K$ in a range from $5.12(24)\times 10^{-12}$ to $9.31(66) \times 10^{-12}$ MeV for kaon masses between 563 and 839 MeV.}

\FullConference{The 30th International Symposium on Lattice Field Theory\\
	June 24-29, 2012\\
Cairns, Australia}

\begin{document}

\section{Introduction}
The $K_L$-$K_S$ mass difference is believed to arise from $K^0$-$\overline{K}^{0}$ mixing via second-order weak interaction. It is highly suppressed and very sensitive to the $V-A$ structure of the weak vertex in the standard model. We proposed a lattice method to compute this mass difference, including both short and long distance effects. Our previous preliminary work on $16^3\times 16 \times 32$ domain wall fermion (DWF) lattice shows promising results ~\cite{Yu:2011gk}. In this work, we extend our calculation to include all operators and various kaon masses on the same lattice. The short distance and long distance contributions will be discussed separately. Then we will discuss the operator mixing and renormalization. Finally, the resulting mass differences will be given in physical units.

\section{Setup of this calculation}

The first-order, $\Delta S=1$ effective weak Hamiltonian including four flavors can be written as:
\begin{equation}
H_W=\frac{G_F}{\sqrt{2}}\sum_{q,q^{\prime}=u,c}V_{qd}V^{*}_{q^{\prime}s}(C_1Q_1^{qq^{\prime}}+C_2Q_2^{qq^{\prime}}),
\label{eq:H_W}
\end{equation}
where the $V_{qq^{\prime}}$ are CKM matrix elements, $C_i$ are Wilson coefficients and we only include the current-current operators, which are defined as:
\begin{equation}
\begin{split}
	Q_1^{qq{\prime}}&=(\bar{s}_iq^{\prime}_j)_{V-A}(\bar{q}_jd_i)_{V-A}\\
 Q_2^{qq{\prime}}&=(\bar{s}_iq^{\prime}_i)_{V-A}(\bar{q}_jd_j)_{V-A}.
\end{split}
\label{eq:operator}
\end{equation}
We neglect the penguin operators in the effective Hamiltonian. This is a good approximation since these operators are suppressed by a factor $\tau = 
|V_{td}V^{*}_{ts}|/|V_{ud}V^{*}_{us}| = 0.0016$ in a four flavor theory.

The essential part of this work is to calculate the four-point correlator: 
\begin{equation}
	\begin{split}
		G(t_f,t_2,t_1,t_i)&=\langle0|T\left\{\overline{K^0}(t_f)H_W(t_2)H_W(t_1)\overline{K^0}(t_i)\right\}|0\rangle \\
										&=N_K^2 e^{-M_K(t_f-t_i)}\sum_{n}\langle\overline{K^0}|H_W|n\rangle\langle n|H_W|K^0\rangle e^{-(E_n-M_K)|t_2-t_1|},
	\end{split}
	\label{eq:unintcorr}
\end{equation}
where $N_K$ is the normalization factor for the kaon interpolating operator, $t_f-t_k$ and $t_k-t_i$ should be sufficiently large to project onto the kaon state. SInce we fix $t_i$ and $t_f$ in the simulation, this correlator depends only on the time separation between the two Hamiltonian $|t_2-t_1|$. We will refer to this quantity as the unintegrated correlator, which receives contributions from all possible intermediate states.

We can integrate the times $t_1$ and $t_2$ in the unintegrated correlator over a time interval $[t_a,t_b]$ and obtain:
\begin{equation}
	\begin{split}
		\mathscr{A}=&\frac{1}{2}\sum_{t_2=t_a}^{t_b}\sum_{t_1=t_a}^{t_b}\langle0|T\left\{\overline{K^0}(t_f)H_W(t_2)H_W(t_1)\overline{K^0}(t_i)\right\}|0\rangle \\ 
            = & N_K^2e^{-M_K(t_f-t_i)} \Bigg\{ \sum_{n\ne n_0}\frac{\langle\overline{K}^0|H_W|n\rangle\langle n|H_W|K^0\rangle}{M_K-E_n}\left( -T + \frac{e^{(M_K-E_n)T}-1}{M_K-E_n}\right) \\
			&+ \frac{1}{2}\langle\overline{K}^0|H_W|n_0\rangle\langle n_0|H_W|K^0\rangle T^2\Bigg\},
	\end{split}
\label{eq:intcorr}
\end{equation}
where $T=t_b-t_a+1$ is the integration range, $|n_0\rangle$ is a state degenerate with kaon if such a state exists. We call this amplitude the integrated correlator. For sufficiently large $T$, the exponentially decreasing terms become negligible. After the subtraction of the exponentially increasing term and a possible quadratic term, the remain term is a linear function of integration range $T$. The coefficient of the linear term gives the finite-volume approximation to $\Delta M_K$ up to some normalization factors. More details about this method can be found in Ref.~\cite{Yu:2011gk}.  

\begin{figure}[!htp]
	\centering
	\begin{tabular}{c|c}
		\hline
		\includegraphics[width=0.35\textwidth]{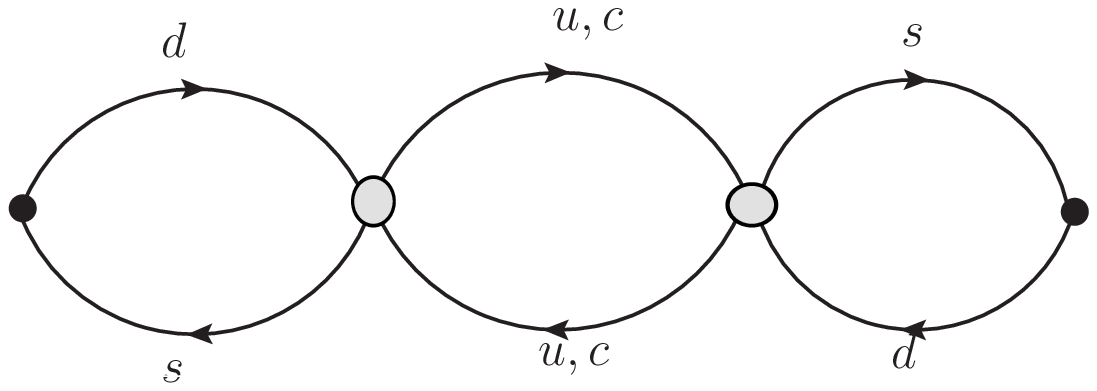} & 
		\includegraphics[width=0.35\textwidth]{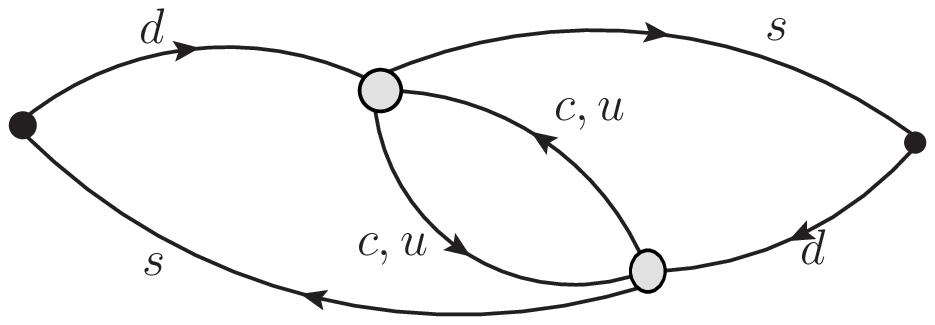} \\
		Type 1 & Type 2 \\
		\hline
		\includegraphics[width=0.35\textwidth]{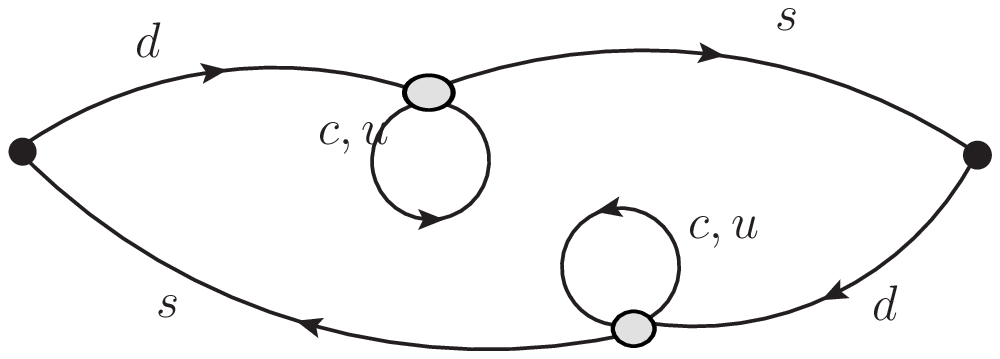} & 
		\includegraphics[width=0.35\textwidth]{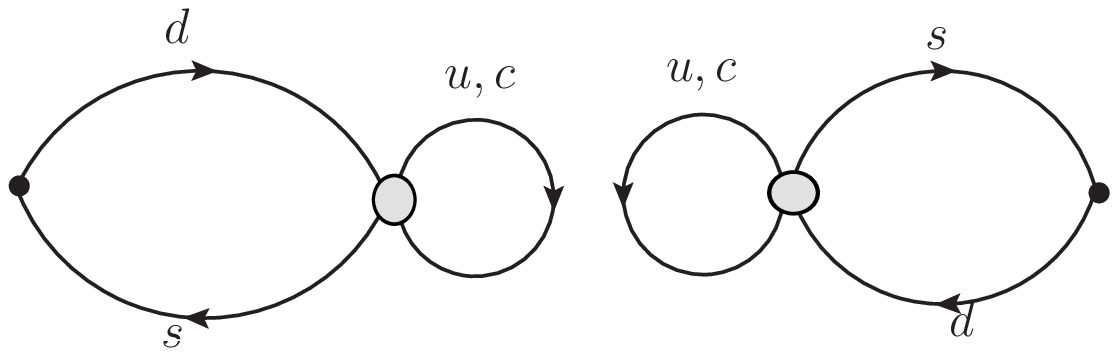}\\
		Type 3, not included & Type 4, not included\\
		\hline
	\end{tabular}
	\caption{Four types of diagrams that contribute to $K^0$-$\overline{K}^0$ mixing. Only type one and type two diagrams are included in this work. The black dots represent a $\gamma^5$ matrix insertion for kaons. the dark circles stand for the four-fermion operators.}
	\label{fig:diagrams}
\end{figure}

There are four types of contractions contributing to this correlator as shown in Fig.~\ref{fig:diagrams}. We only include type 1 and type 2 diagrams in this simulation. This simulation is performed on a $N_f=2+1$ flavors, $16^3\times 32 \times 16$ lattice with domain wall fermions, the Iwasaki action, $a^{-1}=1.73$ Gev, $m_{\pi}=421$ MeV and $m_{K}=563$ MeV. We use 800 configurations, each separated by 10 time units. Two Coulomb gauge wall sources are located at  $t_i=0$ and $t_f=27$. The two weak operators are introduced in the interval $5 \le t_1,t_2 \le 22$. We calculate the four-point function defined in Eq.~\ref{eq:unintcorr} for all possible choices of $t_1$ and $t_2$. The mass difference is obtained from the slope of the integrated correlator plot, and the fitting range is $T\in [9,18]$. 

\section{Short distance contribution}

\begin{figure}[!htp]
\centering
\includegraphics[width=0.5\textwidth]{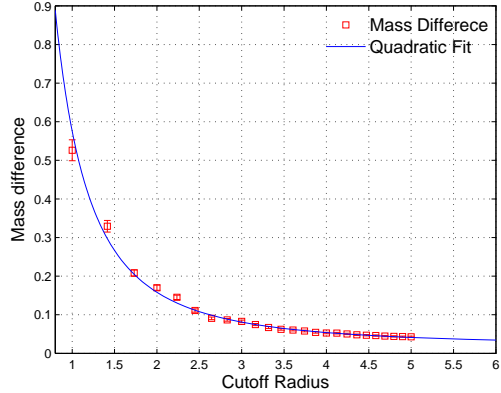}
\caption{The mass difference for different values of the cutoff radius $R$. The blue curve is a naive two-parameter fit to a $1/R^2$ behavior.}
\label{fig:fitcutoff}
\end{figure}

If there is no charm quark in this calculation, the short distance part of $\Delta M_K$ will receive power divergent contributions from heavy intermediate states. Such contributions are unphysical lattice artifacts. To investigate the divergent character of this short distance contribution in detail, we introduce an artificial position-space cutoff radius $R$. When we perform the double integration, we require the space-time separation between the positions of the two operators to be larger than or equal to this cutoff radius :   
\begin{equation}
\sqrt{(t_2-t_1)^2+(\vec{x}_2-\vec{x}_1)^2}\geq R.
\end{equation}
We can plot the mass difference $\Delta M_K$ as a function of this cutoff radius $R$. We use 600 configurations separated by 10 time units, with $m_\pi=412$ MeV and $m_K=563$ MeV. We only include the operator combination $Q_1\cdot Q_1$, {\it i.e.} both four quark operators are $Q_1$ operators. In Fig.~\ref{fig:fitcutoff}, we show the cutoff dependence of mass difference. The blue curve is a naive uncorrelated two parameter fit:
\begin{equation}
\Delta M_K^{11} (R)= \frac{b}{R^2} + c,
\end{equation}
where $b$ and $c$ are constants. The fitting results shows a power divergent short distance contribution.

We introduce a valence charm quark to control such unphysical short distance effects. In our earlier work~\cite{Yu:2011gk}, we found that GIM mechanism substantially reduce the short distance contribution. One might expect that the GIM cancellation would leave us a logarithmic divergence of form $\mathrm{ln}(m_c a)$ on lattice. However, because of the $V-A$ structure of the weak vertices, the GIM cancellation is complete, leaving only convergent pieces. Thus if the lattice artifacts associated with large $m_ca$ and the quenched treatment of the charm quark can be neglected, our four flavors calculation should capture all important aspects of $\Delta M_K$. 

In Ref.~\cite{Yu:2011gk}, we were unaware of the absence of a $\mathrm{ln}(m_c a)$ term in the lattice calculation with valence charm and we performed an explicit Rome-Southampton style, RI/SMOM subtraction to remove it. In fact, the subtraction term reported in Ref.~\cite{Yu:2011gk}, performed at a scale $\mu=2$ GeV, was zero within errors, consistent with the absence of a true short distance contribution to  $\Delta M_K$.

\section{Long distance contribution}
In this section we will discuss the long distance contribution to our calculation of $\Delta M_K$. The integrated correlator receives contribution from both short and long distances. Therefore, in this section we examine the unintegrated correlators in Eq~.\ref{eq:unintcorr}, which depends only on the time separation $T$ of two effective Hamiltonians. If the separation $T$ is sufficiently large, the contribution from the lightest intermediate state will dominate the signal. There is no vacuum state since we don't include disconnected diagrams, so the two lightest states are $\pi^0$ and $\pi$-$\pi$ states. These two states have different parity. We can examine them separately by separating each left-left $\Delta S=1$ operator into parity conserving and violating parts. The results presented in this section are for an average over 800 configurations separated by 10 time units, with a valence light quark mass $m_l=421$ MeV and various strange quark masses.

\begin{figure}[!htp]
	\centering
\begin{tabular}{cc}
\includegraphics[width=0.4\textwidth]{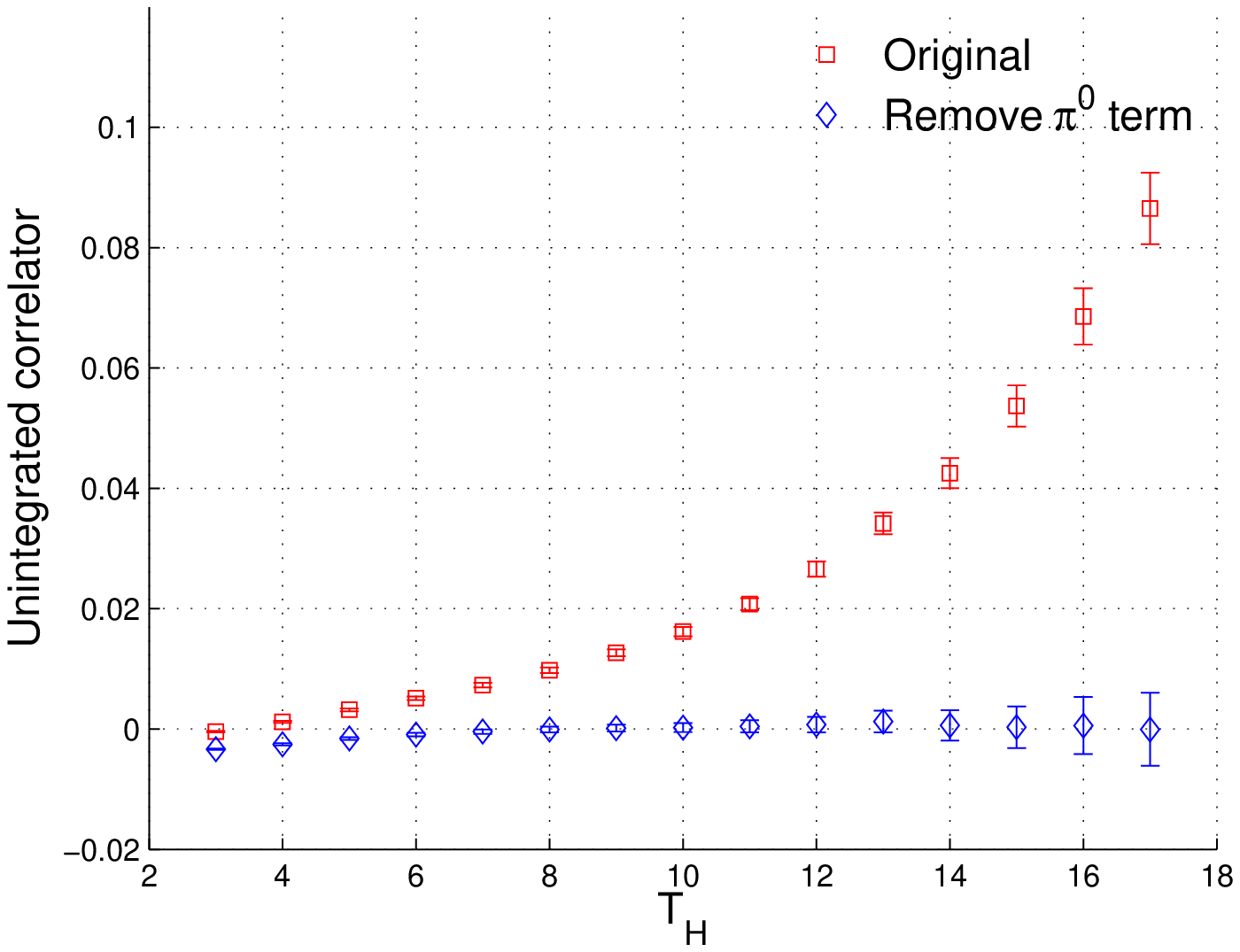} &
\includegraphics[width=0.4\textwidth]{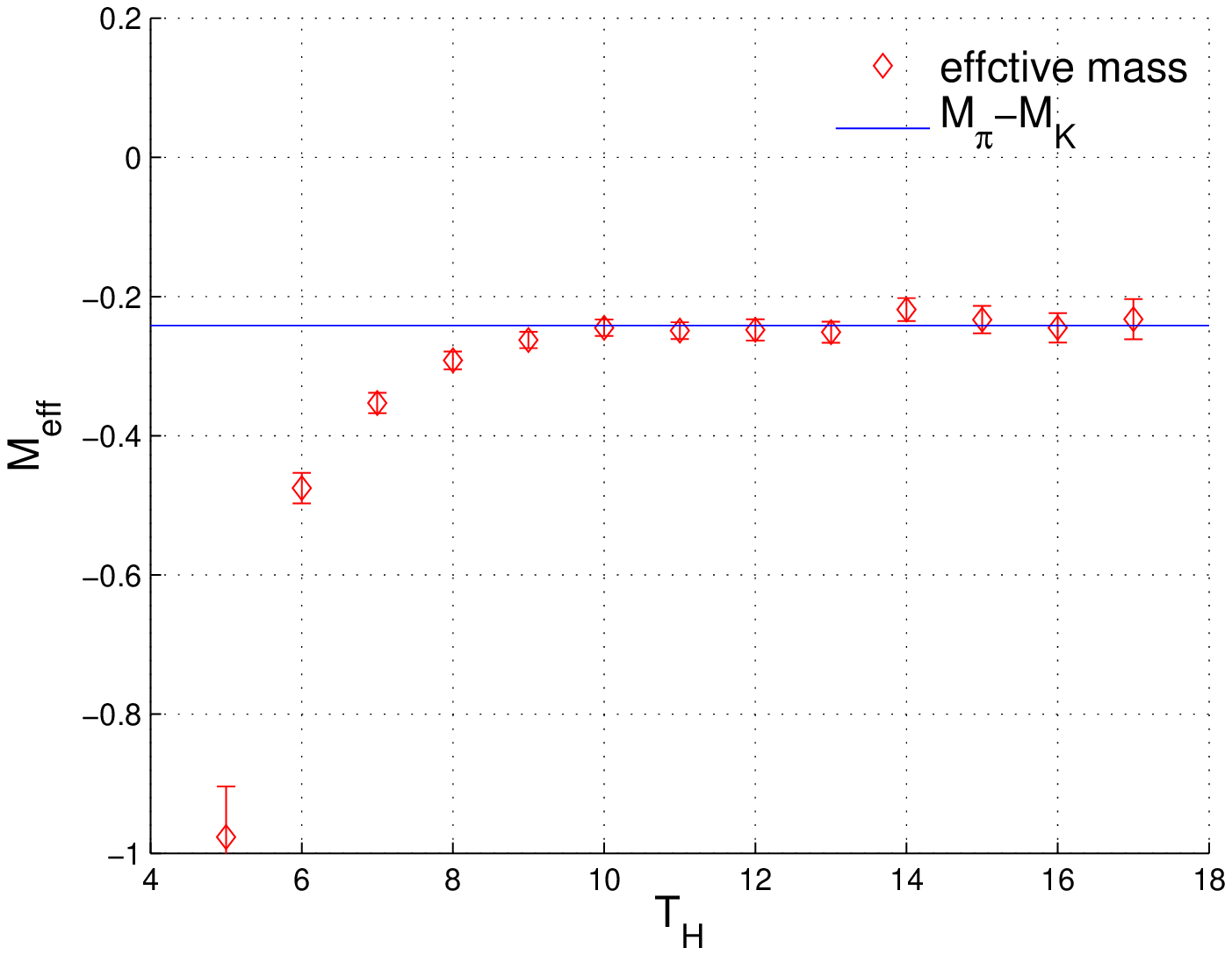} 
\end{tabular}
\caption{A plot of the unintegrated correlator and resulting effective mass for the parity-conserving operator combination  $Q_1 \cdot Q_1$ and a kaon mass $_K=839$ MeV. In the left-hand plot, the red diamonds and blue squares show the result before and after subtraction of the $\pi^0$ term. In the right-hand plot, the red diamonds are effective masses obtained from the unintegrated correlator. The blue horizontal line shows the ``exact'' value of $M_{\pi}-M_K$ obtained from the two point correlator calculation.}
\label{fig:q1q1unintcorr}
\end{figure}

We will discuss the parity-odd channel first. For this case, both weak operators are parity conserving which implies that all intermediate states have odd parity. In Fig.\ref{fig:q1q1unintcorr}, we plot the unintegrated correlator from operator combination $Q_1\cdot Q_1$ and the resulting effective mass for the choice of kaon mass $M_K=0.4848(8)$. In the plots of the unintegrated correlators we show both original results and the results after the subtraction of the $\pi^0$ contribution. This subtraction is done using the $\langle \pi^0 | Q_i | K^0 \rangle$ matrix element from a three point correlator calculation. Since only the $\pi^0$ term should be present for large time separations, we expect that the results after subtraction should be consistent with zero for large $T_H$.  In the effective mass plots, we calculate the effective mass $M_X-M_K$ from the unintegrated correlators, here $M_X$ is the mass of the intermediate state. For this parity conserving case, the lightest state is the pion. The ``exact'' $M_{\pi}-M_K$ mass obtained from the two point correlator calculation is shown in the plots as a blue horizontal line which agrees well with the computed effective mass.  

\begin{figure}[!htp]
	\centering
\begin{tabular}{cc}
$Q_1 \cdot Q_1$ & $Q_1 \cdot Q_2$ \\
\includegraphics[width=0.4\textwidth]{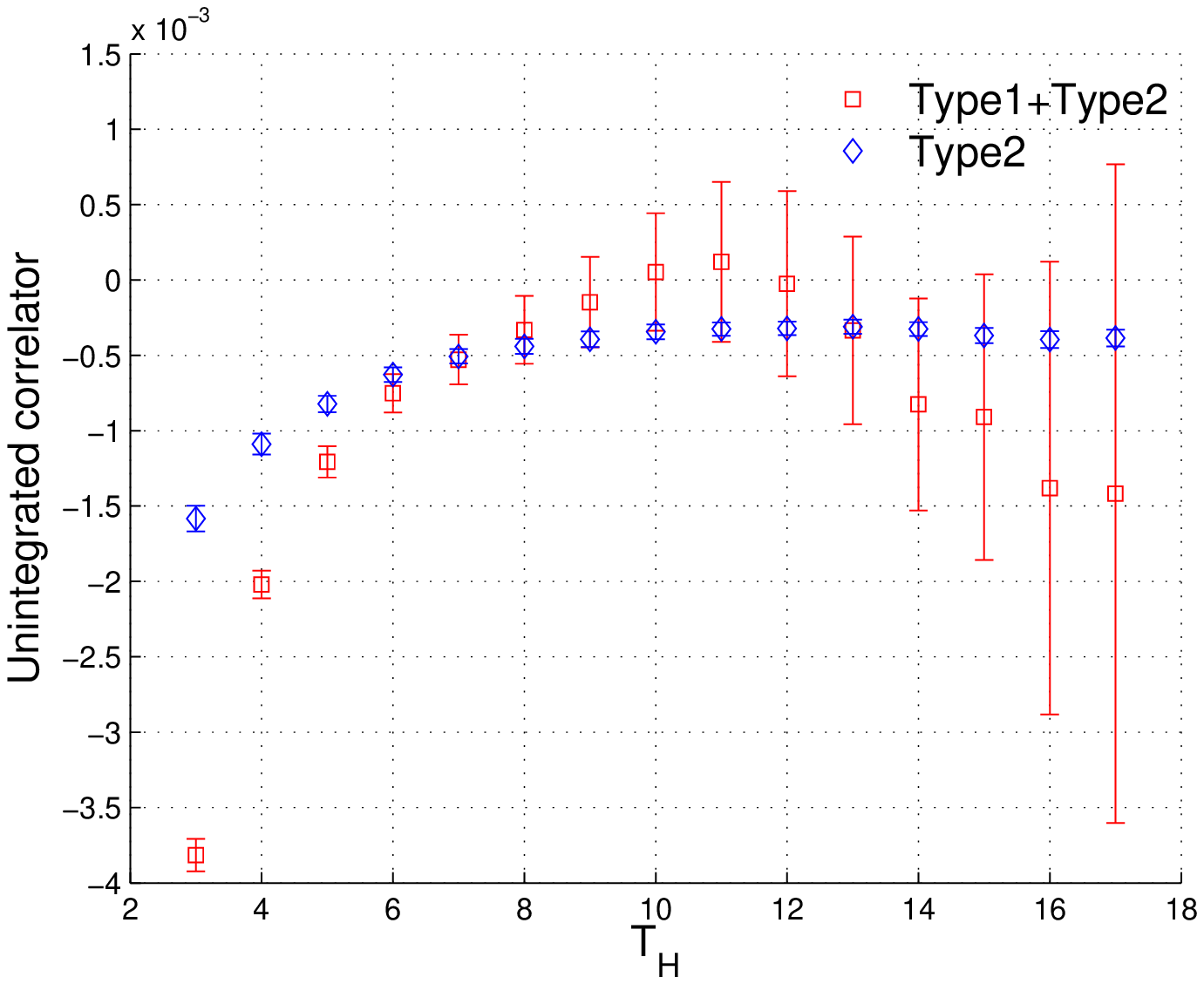}&
\includegraphics[width=0.4\textwidth]{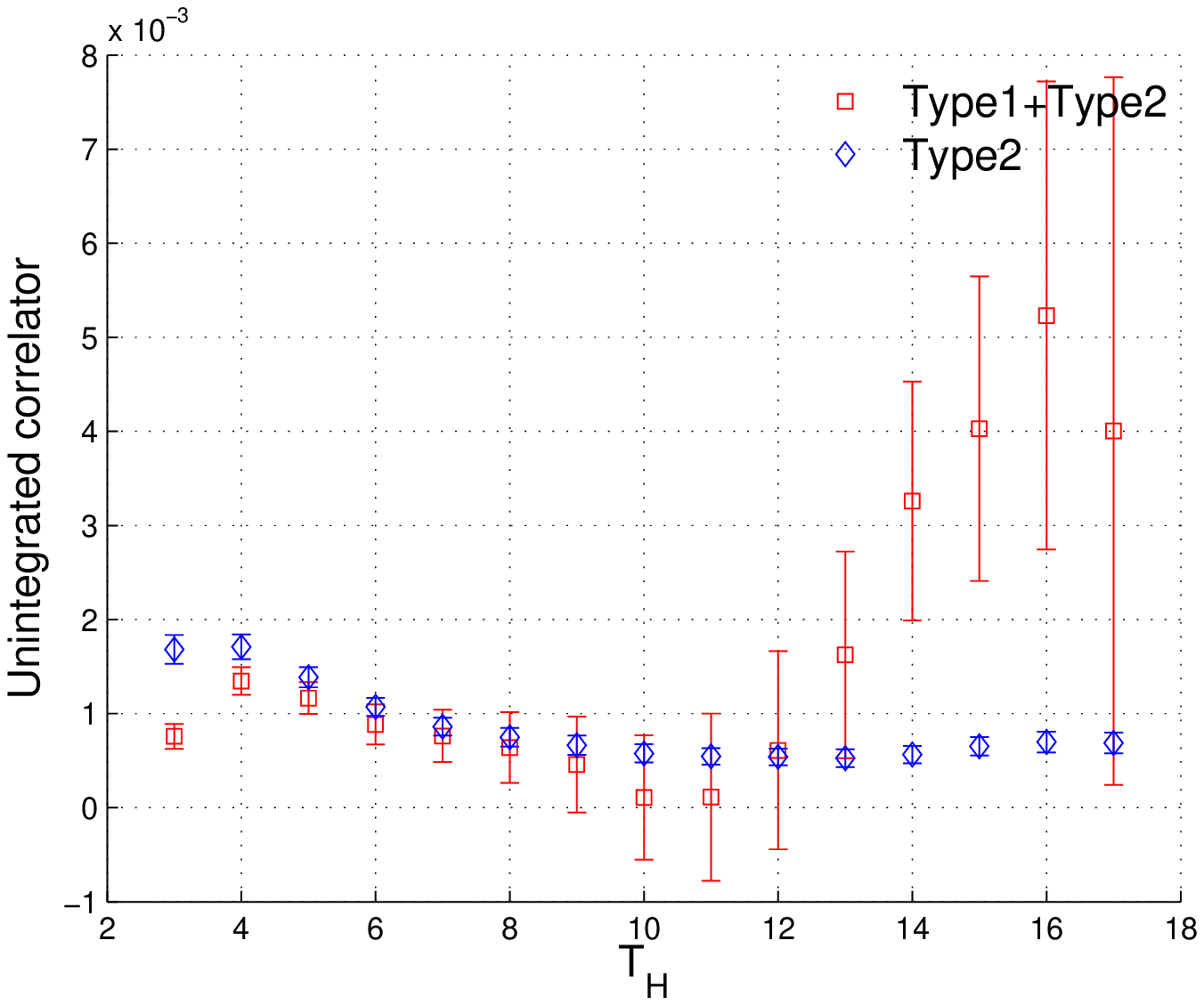}\\
$Q_2 \cdot Q_2$ & Intermediate state mass\\
\includegraphics[width=0.4\textwidth]{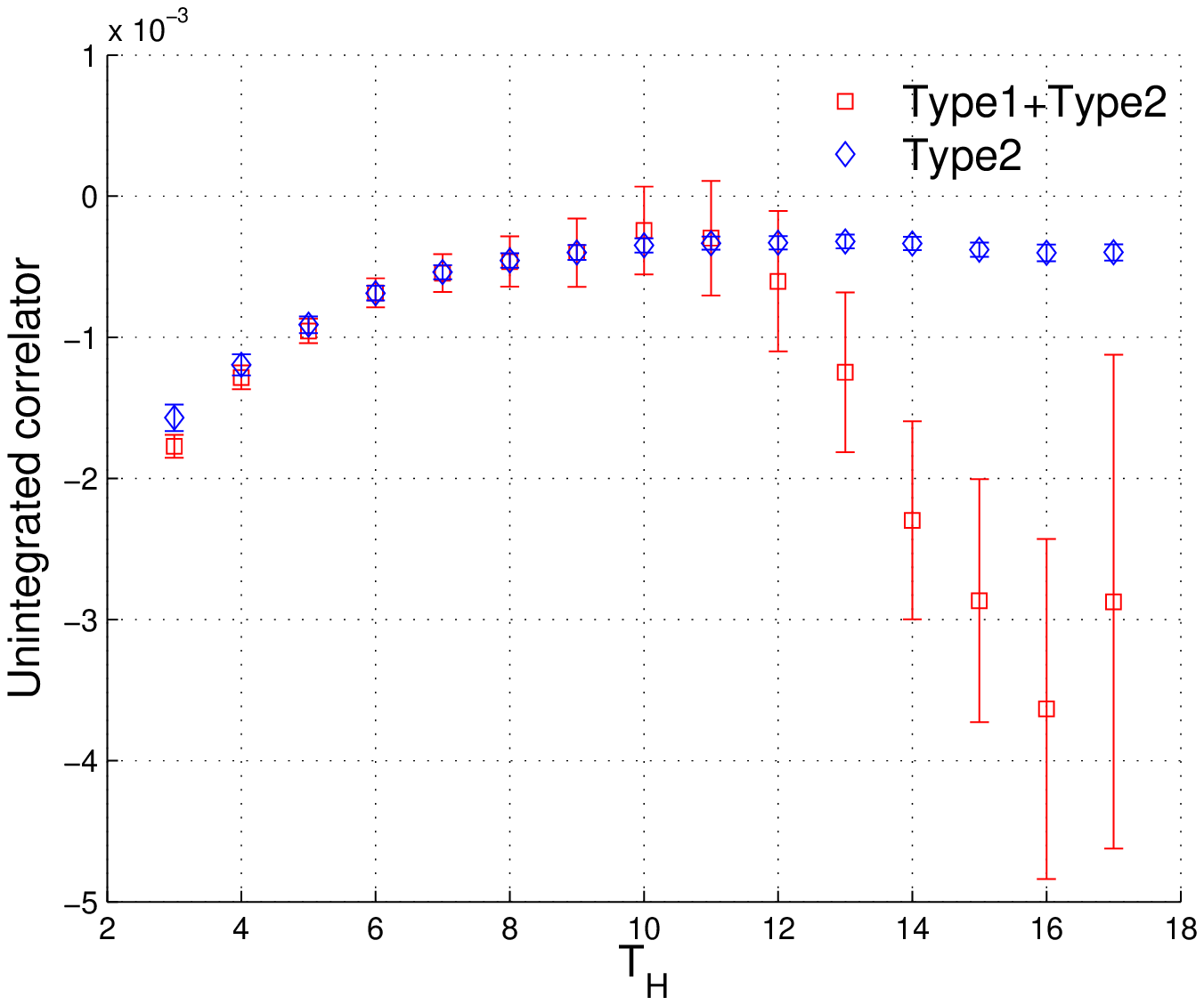}&
\includegraphics[width=0.4\textwidth]{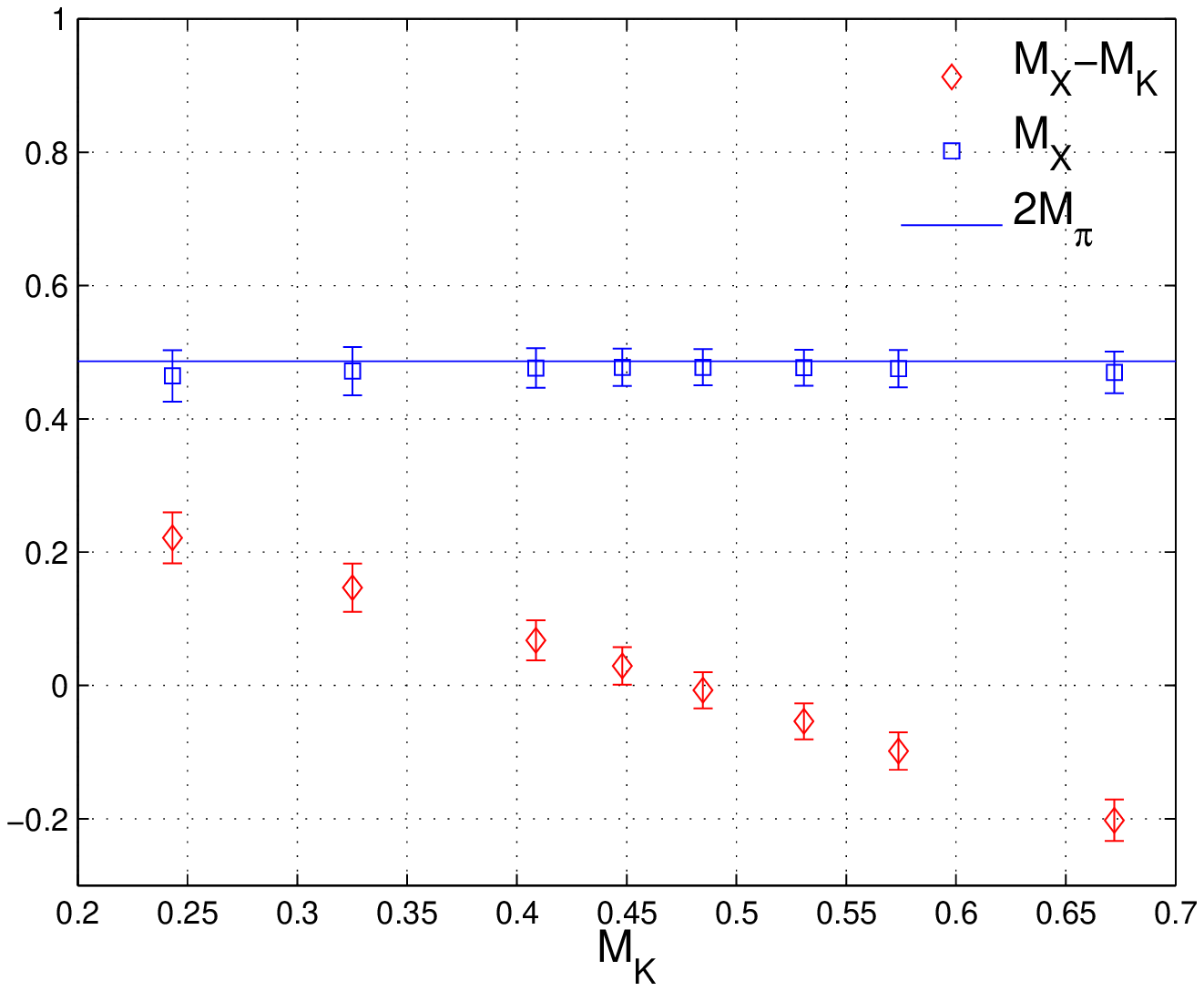}
\end{tabular}
\caption{The unintegrated correlators for different, parity-violating operator combinations at a kaon mass $M_K=839$ MeV. We plot both the full results and the results from type 2 diagrams only. The last plot is the fitted intermediate state mass (red diamonds) and the sum of that mass and the kaon mass (blue squares) for all choices of kaon masses. The results shown in this last plot are obtained from fitting the type 2 diagrams alone. }
\label{fig:pvunintcorr}
\end{figure}

Next we will examine the case where parity violating operators appear at both vertices, so the intermediate states can only be parity even. In Fig.~\ref{fig:pvunintcorr}, we present the unintegrated correlators for the three different operator combinations evaluated at a kaon mass $M_K=839$ MeV. This kaon mass is very close to the energy of two pions at rest, so we expect to find a non-zero plateau at large time separation $T_H$.  However, our results are extremely noisy at long distance and we are not able to identify such a plateau. The large noise can be explained as follows.  Although the signal should come from two-pion intermediate states, we will also have noise, whose large time behavior is like that of a single pion intermediate state. Then the signal to noise ratio will fall exponentially for large time separation. The situation here is very similar to what is found for disconnected diagrams. We also expect that most of the noise comes from type 1 diagrams, shown in Fig.~\ref{fig:diagrams}, because the topology of type 2 diagrams does not allow a single-pion contribution to their noise. This argument is confirmed by plotting the results from type 2 contractions only.  If we analyze the type 2 diagrams alone, and fit the resulting intermediate state masses the results agree with the two-pion mass very well, as seen in the lower right panel of Fig.~\ref{fig:pvunintcorr}.

\section{The $K_L$-$K_S$ mass difference}

In order to calculate the physical  $K_L-K_S$ mass difference, we must connect our four-quark lattice operators with the physical $\Delta S=1$ effective weak Hamiltonian $H_W$. Thus, we must determine the Wilson coefficients and normalize the lattice operators in the same scheme in which the Wilson coefficients are computed. This procedure can be summarized by : 
\begin{equation}
\begin{split}
H_W&=\frac{G_F}{\sqrt{2}}\sum_{q,q^{\prime}=u,c}V_{qd}V^{*}_{q^{\prime}s}\sum_{i=1,2}C_i^{\overline{\rm MS}}(\mu)(1+\Delta r^{{\rm RI}\rightarrow\overline{\rm MS}})_{ij}(Z^{lat\rightarrow RI})_{jk}Q_k^{qq^{\prime},{\rm lat}}(\mu)\\
&=\frac{G_F}{\sqrt{2}}\sum_{q,q^{\prime}=u,c}V_{qd}V^{*}_{q^{\prime}s}\sum_{i=1,2}C_i^{\rm lat}(\mu)Q_i^{qq^{\prime},{\rm lat}}(\mu).
\end{split}
\end{equation}
All the operator renormalization and mixing are performed at a scale $\mu=2.15$ GeV. The Wilson coefficients $C_i^{\overline{\rm MS}}(\mu)$ are calculated following the formulas in Ref.~\cite{Buchalla:1995vs}. The matching matrix $\Delta r^{{\rm RI} \to \overline{\rm MS}}$ is provided by Christoph Lehner. The lattice operator mixing matrix is obtain from a non-perturbative renormalization calculation~\cite{Blum:2001xb}. Combining all three ingredients we can obtain the coefficients $C^{\rm lat}_i$ for the bare lattice operators.  These results are given in Tab.~\ref{tab:npr}.

\begin{table}[!htp]
\caption{The Wilson coefficients, the ${\rm RI} \to \overline{\rm MS}$ matching matrix, the non-perturbative ${\rm lat} \to {\rm RI}$ operator renormalization matrix and their final product, all at a scale $\mu=2.15$ GeV shown in columns one through four respectively.}
\begin{tabular}{cc|cc|cc|cc}
	\hline
$C_1^{\overline{\rm MS}}$ & $C_2^{\overline{\rm MS}}$ & $\Delta r_{11} = \Delta r_{22}$ &  $\Delta r_{12} = \Delta r_{21}$ & $Z_{11}=Z_{22}$ & $Z_{12}=Z_{21}$ & $C_1^{\rm lat}$ & $C_2^{\rm lat}$ \\
\hline
-0.2976 & 1.1391 & -$6.562\times 10^{-2}$ & $7.521\times 10^{-3}$ & 0.5725 & -0.01412 & -0.1693 & 0.6119 \\ 
\hline
\end{tabular}
\label{tab:npr}
\end{table}

We now combine all these ingredients and determine the mass difference $\Delta M_K$ in physical units. We present results for four kaon masses ranging from 563 MeV to 839 MeV. In the heavy kaon cases, especially the $839$ MeV kaon, the two-pion intermediate state will be close to being degenerate with the kaon. Thus,we should remove the $T^2$ term in Equation~\ref{eq:intcorr} arising from this degenerate $\pi-\pi$ state. However, we are not able to identify the two-pion intermediate state within errors. This implies that this on-shell, two-pion intermediate state contributes only a small part to the mass difference in our calculation.  Therefore, for all the results presented in this section, we neglect possible $T^2$ contamination from a degenerate two-pion intermediate state.

\begin{table}[!htp]
\caption{The contribution of the three operator products evaluated here to the mass difference $\Delta M_K$ for four different choices of the kaon mass. The final column gives the complete long and short distance contribution to $\Delta M_K$ expressed in physical units.}
\centering
\begin{tabular}{c|ccc|c}
\hline
	$M_K$ (MeV)     & $\Delta M_K^{11}$ 
& $\Delta M_K^{12}$ 
& $\Delta M_K^{22}$
& $\Delta M_{K}$ ($\times 10^{-12}$ MeV) \\
\hline
563 & 6.42(15)  & -2.77(16)   & 1.56(9)    & 5.12(24)\\
707 & 8.94(23) & -3.16(27)   & 2.26(14)  & 6.92(39)\\
775 & 10.65(29) &  -3.49(35)  & 2.67(18)  & 8.08(51)\\
839 & 12.55(37) &  -3.84(46)  & 3.11(24)  & 9.31(66)\\
\hline
\end{tabular}
\label{tab:mass}
\end{table}

In summary, we perform a systematic study of the $K_L-K_S$ mass difference calculation using the method of lattice QCD. We only include part of the contractions and use a pion mass of $421$ MeV. Our results range from $5.12(24)\times 10^{-12}$ MeV to $9.31(66)\times 10^{-12}$ MeV for kaon masses between $563$ MeV and $839$ MeV, while the physical result is $3.48\times 10^{-12}$ MeV. 

The author thank very much all my colleagues in the RBC and UKQCD collaborations for valuable discussions and suggestions. Especially thanks to Prof. Norman Christ for detailed instructions and discussions.

\bibliography{lattice2012yujl}
\bibliographystyle{JHEP}

\end{document}